\documentstyle[preprint,aps]{revtex}

\begin{document}
\bibliographystyle{prsty}
\draft
 
\title{Quantum-State Engineering of Multiple Trapped Ions for
Center-of-Mass Mode
\thanks{This work was supported by the National Natural Science Foundation of
China under Grant No.19734006 and Chinese Academy of Science}}
\author{Hao-Sheng Zeng$^{1,2}$, Xi-Wen Zhu$^1$ and Ke-Lin Gao$^1$} 
\address{$^1$Laboratory of Magnetic Resonance and Atomic and Molecular
Physics, Wuhan Institute of Physics and Mathematics, Chinese Academy of
Science, Wuhan 430071, People's Republic of China \\
$^2$Department of Physics, Hunan Normal University, Hunan
410081, People's Republic of China}
 
\maketitle

\begin{abstract}
We propose a scheme to generate a superposition with
arbitrary coefficients on a line in phase space for the center-of-mass
vibrational mode of $N$ ions by means of isolating all other spectator
vibrational modes from the center-of-mass mode. It can be viewed as the
generation of previous methods for preparing motional states of one ion. For
large number of ions we need only one cyclic operation to generate such a
superposition of many coherent states.
\end{abstract}
\vskip 1cm

\narrowtext

\begin{center}
{\bf I. INTRODUCTION}
\end{center}

In recent years, there has been much interest in the preparation of
nonclassical quantum states, especially the engineering of arbitrary quantum
states, in order to study fundamental properties of quantum mechanics. Many
schemes have been proposed for the purpose of engineering various quantum
states[1-8], especially the superpositions of coherent states on a circle or
superpositions of coherent states on a line in the field of cavity and
trapped ions[9-12]. Because discrete superpositions of coherent states on a
circle or on a line may approximate many quantum states[9], such as number
states, amplitude-squeezed states and quadrature squeezed states, which
provides a new way for quantum-state engineering.

In the previous papers on ion trap, one usually prepares the center-of-mass
vibrational mode (COM mode) of one trapped ion to some quantum states, where
only one vibrational mode has been involved. In this paper we shall consider
the general case of $N$ ions, where many other spectator vibrational modes
involve except for COM mode. We find that in our method one can still
prepares the COM mode to a discrete superposition of coherent states on a
line in phase space, and the spectator vibrational modes do not affect the
quantum-state engineering of the COM mode. We also find that for large
number of ions one needs only one cyclic operation to prepare a
superposition of a number of coherent states of COM mode on a line which may
need many cycles for one ion. Thus the operation time for quantum state
preparation can be greatly reduced, which is of importance for the
experimental realization in the view of decoherence.

As the usual treatment for quantum computation, we employ the ground state $%
\left| 0\right\rangle $ and a particular long-lived metastable excited state 
$\left| 1\right\rangle $ of the trapped ion. The transition between these
two internal states is dipole forbidden, but we can drive it by irradiating
the ion with a pulsed laser that tuned to the frequency of the transition.
In order to measuring the internal stated, we then employ the third
auxiliary electric level $\left| 2\right\rangle $ of the ion[5,13,14] which
is a strong electric transition to the ground state $\left| 0\right\rangle $%
. By directing a laser resonant with the transition $\left| 0\right\rangle
\rightarrow \left| 2\right\rangle $ on the ion, and then probing for the
occurrence of fluorescence light, the electric quantum state is projected
either into the ground or excited state, conditioned on the observation of
fluorescence or of no-fluorescence event. Since any spontaneously emitted
photon will disturb the motional quantum state via recoil effects, so we
consider only those events where no fluorescence have been observed, i.e.
the case that the internal state is projected into the excited state $\left|
1\right\rangle $.

This paper is organized as follows: In Sec.II{\rm \ }we introduce the
laser-atom interaction mode used for the engineering scheme, which was first
proposed by S$\phi $rensen and M$\phi $lmer. In Sec.III we describe the
method to generate arbitrary superpositions of coherent states of COM mode
on a line in phase space for $N$ ions in Lamb-Dicke limit and low-excitation
regime, where we assumed that the two lasers performing on each ion have
very small detunings to the red and blue sidebands for COM mode, so that we
can omit the effects from all other spectator vibrational modes for a while.
And in Sec.IV, we then discuss the case of the existence of all other
vibrational modes to demonstrate the validity of the conclusion deduced from
the previous section. A summary is given in Sec.V.

\begin{center}
{\bf II. LASER-ATOM INTERACTION}
\end{center}

According to the consideration of S$\phi $rensen and M$\phi $lmer[15,16], we
consider a string of $N$ ions trapped in a linear trap, which are strongly
bounded in the $y$ and $z$ direction but weakly bounded in a harmonic
potential in $x$ direction. Assuming that every ion was illuminated
simultaneously with two lasers, then the Hamiltonian describing the
interaction between the system and lasers is

\begin{equation}
H=H_0+H_{int},  
\end{equation}
$$
H_0=\nu (a^{+}a+1/2)+\omega \sum_i^N\sigma _{iz}, 
$$
$$
H_{int}=\sum_i^N\frac \Omega 2\left\{ \sigma _{+i}\left[ e^{-i\omega
_1t}+e^{-i\omega _2t}\right] e^{i\eta (a+a^{+})}+H.C.\right\} 
$$

where $\nu$ is the frequency of the vibration, $a^+$ and $a$ are the ladder
operators of the quantized oscillator, $\omega $ is the energy difference
between the ground state $\left| 0\right\rangle $ and the long-lived
metastable excited state $\left| 1\right\rangle $ of the ion. $\omega
_1=\omega +\delta ~$ and $\omega _2=\omega -\delta ,$ are the frequencies of
the two lasers addressing on each ion, and $\delta $ is the detuning. We
have assumed that the lasers are very close to the blue and red sidebands $%
\left( \delta \approx \nu \right) ,$ so that we can neglect the contribution
from all other vibrational modes for a while and consider only the one from
the COM mode. We also assumed that all ions participate equally in the
vibration, so that the coupling of the recoil to the vibration is identical
for all ions, i.e., $\eta _i=\eta $ for all $i.$ For simplicity we still
more assume that the Rabi frequency for all ions participating in the
operation is same. In this paper we will consider the behavior of the ions
in the Lamb-Dicke regime,$\eta \ll 1.$ In this limit we can expand the
Hamiltonian up to the first order in $\eta .$ In the interaction picture
with respect to $H_0$, it becomes

\begin{equation}
H_{int}=4\Omega J_x\cos \delta t-2\sqrt{2}\eta \Omega J_y[x(\cos (\nu
-\delta )t+\cos (\nu +\delta )t)+p(\sin (\nu -\delta )t+\sin (\nu +\delta)t)],  
\end{equation}
where we have introduced the dimensionless position and momentum operators, $%
x=(1/\sqrt{2})(a+a^{+})$, and $p=(i/\sqrt{2})(a^{+}-a)$, and the collective
spin operators $J_\alpha =\sum_i^Nj_{\alpha i}=\sum_i^N\sigma _{\alpha i}/2$ 
$(\alpha =x,~y,~z)$.

Choosing laser intensities $\Omega \ll \delta ,$and tuning close to the
sidebands $\nu -\delta \ll \delta ,$ we may neglect the $J_x$ term and the
terms oscillating at frequency $\nu +\delta $, and the Hamiltonian reads,

\begin{equation}
H_{int}=[f(t)x+g(t)p]J_y  
\end{equation}

with $f(t)=-2\sqrt{2}\eta \Omega \cos \left( \nu -\delta \right) t$ and $%
g(t)=-2\sqrt{2}\eta \Omega \sin \left( \nu -\delta \right) t$. Note that the
spin operators for different ions commute, so the propagator for this
Hamiltonian is,

\begin{equation}
U=\prod_{i=1}^N \frac 12\left\{ [D^{+}(\beta )+D(\beta )]-\sigma
_{yi}[D^{+}(\beta )-D(\beta )]\right\} 
\end{equation}

where $\beta =i\eta \Omega te^{i(\nu -\delta )t}$ and $D(\beta )=e^{\beta
a^{+}-\beta ^{*}a}$.

\begin{center}
{\bf III. QUANTUM-STATE ENGINEERING FOR COM MODE}
\end{center}

In this section we start to discuss the quantum-state engineering for COM
mode. Let us neglect the effects from all other vibrational modes for a
while and consider only the contribution from COM mode, so that we can
employ the Hamiltonian or propagator discussed in the previous section.

{\bf 1. The case for one ion(N=1).}

We first discuss a simple case where only one ion is contained in the
system. In this case the system only has one motional mode---COM mode. We
assume the initial internal state of the trapped ion to be in an arbitrary
superposition of ground state $\left| 0\right\rangle $ and metastable
excited state $\left| 1\right\rangle $ which can be realized by the
application of a laser field resonant with the electric transition, and the
motional state to be in coherent state $\left| \alpha \right\rangle $, so
that the whole initial state of the ion is,

\begin{equation}
\left| \Psi (t_0)\right\rangle =\frac 1{\sqrt{1+\left| p_1\right| ^2}}\left(
\left| 1\right\rangle +ip_1\left| 0\right\rangle \right) \left| \alpha
\right\rangle  
\end{equation}
where the parameter $p_1$ controls the weights of the two electric level, $i$
denotes the imaginary unit.

We now illuminate the atom with two lasers discussed above, i.e., performing
the propagator (4) with $N=1$ on the initial state (5), for an interaction
time $\Delta t$ (For convenience we denote $\Delta t=t $ in the expression 
of displacing factor $\beta $), following by a measurement on the electric state. With no
fluorescence has been detected, then the internal quantum state is projected
into the excited state $\left| 1\right\rangle $, and the resulting
conditioned vibrational quantum state at time $t_1$ reads,

\begin{equation}
\left| \Psi (t_1)\right\rangle =\frac 1{2\sqrt{1+\left| p_1\right| ^2}%
}\left[ (1-p_1)D(\beta )+(1+p_1)D(-\beta )\right] \left| 1\right\rangle
\left| \alpha \right\rangle 
\end{equation}

By appropriate resonant laser interaction we now set the electric state to
again be in a superposition of type (5) but with weighting factor $p_2$, and
realize a second identical interaction with laser field and subsequent
probing for fluorescence. For the case that again no fluorescence will be
detected, one obtain the conditional vibration state,

\begin{equation}
\left| \Psi (t_2)\right\rangle =\aleph _2\prod_{i=1}^2 \left[ (1-p_i)D(\beta
)+(1+p_i)D(-\beta )\right] \left| 1\right\rangle \left| \alpha \right\rangle 
\end{equation}

with $\aleph _2=\frac 14\left[ (1+\left| p_1\right| ^2)(1+\left| p_2\right|
^2)\right] ^{-1/2}$.

If we repeat the above procedure $n$ cycles with each cycle having the same
interaction time $\Delta t$, following by a detection of no fluorescence,
then the motional state for COM mode is,

$$
\left| \Psi (t_n)\right\rangle =\aleph _n\prod_{i=1}^n\left[ (1-p_i)D(\beta
)+(1+p_i)D(-\beta )\right] \left| \alpha \right\rangle 
$$
\begin{equation}
=\aleph _n\sum_{k=0}^nC_n^kD\left[ (2k-n)\beta \right] \left| \alpha
\right\rangle  
\end{equation}

with $\aleph _n=\prod_{i=1}^n\frac 12(1+\left| p_i\right| ^2)^{-1/2}$. and

\begin{equation}
C_n^k=\sum_{\left\{ k\right\} }\prod_{i\in \left\{ k\right\}
}(1-p_i)\prod_{j\in \left\{ n-k\right\} }(1+p_j)  
\end{equation}

where $\left\{ k\right\} $ denote a set that picking $k$ numbers out of $n$
natural numbers corresponding to $n$ parameters $p_i$, and $\left\{
n-k\right\} $ denote the complementary set to $\left\{ k\right\} $. The
first $\prod $ represents the multiplication of $k$ factors $\left(
1-p_i\right) $ with $i$ being the element of set $\left\{ k\right\} $, and
the $\prod $ represents the multiplication of $\left( n-k\right) $ factors $%
\left( 1+p_j\right) $ with $j$ being the element of set $\left\{ n-k\right\} 
$. The sum at there is for all possible different sets $\left\{ k\right\} $
that formed by picking $k$ numbers out of $n$ natural numbers.

(8) represents a discrete superposition of coherent states on a line in
phase space with distance $2\beta $ between them and centered around the
phase-space point of the initial coherent state $\left| \alpha \right\rangle 
$. For the purpose of state-preparation, the coefficients $C_n^k(k=0,\cdots
,n)$ are usually known, so our task is how to determine those electronic
state parameters $p_{i~}(i=1,\cdots ,n)$ from known coefficients $C_n^k$.
Note that (8) is unnormalized and the independent numbers of $C_n^k$ is $n$
which equals to the numbers of $p_i$, so we can determine all parameters $%
p_{i~}$uniquely from (9) in combination with the normalized condition. But
generally one may have several realizations for parameters $p_i$, and one
should choose that realization where the probability of the sequence of
no-fluorescence events,

\begin{equation}
P_n=\left( \frac 14\right) ^n\prod_{i=1}^n\frac 1{1+\left| p_i\right| ^2}%
\end{equation}
is maximal (i.e.$\left| p_i\right| \sim 1$). Also note that if we expect to
obtain all parameters $p_i$ from (9), it will be very difficult when $n$ is
large. Fortunately we have another method to solve this problem. similar to
the reference[12], we can deduce the following recurrence relation for
coefficients $C_n^k$,

\begin{equation}
C_n^k=(1+p_n)C_{n-1}^k+(1-p_n)C_{n-1}^{k-1} 
\end{equation}

with $C_{n-1}^{-1}=C_{n-1}^n=0$, and equations for $x_n~$($%
x_n=(p_n-1)/(p_n+1)$),

\begin{equation}
\sum_{k=0}^nC_n^{n-k}(x_n)^k=0.  
\end{equation}

When all coefficients $C_n^k$ are known, we can firstly solve out $x_n$ (or $%
p_n$) from (12) and then substitute it into (11) to find previous
coefficients $C_{n-1}^k$. With all coefficients $C_{n-1}^k$ known, we can
obtain $p_{n-1}$ and coefficients $C_{n-2}^k$ by repeating the same
procedure. And so on, until we have obtained all parameters $p_i$ for the
preparation of internal state of the ion. Thus we successfully prepared a
superposition of coherent states on a line with required coefficients for
the motion of single trapped ion.

It is worthwhile to note that the superpositional form (8) of coherent
states for COM mode is the same as the result of reference [12], but the
displacing factor $\beta$ is different. In reference [12], the ion is driven
by two lasers resonant with the well resolved upper and lower vibrational
sidebands. But in our method, the detuning $\delta$ is not equal to the
frequency of COM vibrational mode, but very close to it, so that the phase
of the displacing factor changes very slowly over time. Obviously, when $%
\delta =\nu$, the two results are consistent completely.

{\bf 2. The case for two ions (N=2).}

This case is usually viewed as the typical candidate used to implement the
universal two-bit gate for quantum computation in an ion trap. In this case,
the two ions are confined symmetrically in the two sides of the center of
the linear trap along the axis. The external confining potential is
symmetric about the center of the trap. Due to the Coulomb repulsive force
between them, the two ions retain a certain distance. Similar to the above
discussion, we prepare the initial state of the system of the two ions to be,

\begin{equation}
\left| \Psi (t_0)\right\rangle =\aleph _0\left( \left| 1\right\rangle
_1+ip_1\left| 0\right\rangle _1\right) \left( \left| 1\right\rangle
_2+ip_2\left| 0\right\rangle _2\right) \left| \alpha \right\rangle 
\end{equation}

where $\aleph _0=\left[ (1+\left| p_1\right| ^2)(1+\left| p_2\right|
^2)\right] ^{-1/2}$ with parameters $p_1 $ and $p_2 $ being the controlling
weights of the electronic levels of the two ions respectively. Then we
illuminate both the two ions simultaneously with two lasers , i.e.,
performing the propagator (4) with $N=2$ on the above initial state (13),
for an interaction time $\Delta t$, following by a measurement on the
internal state of two ions. If we detect the two ions both in excited state $%
\left| 1\right\rangle $, then the conditioned state for the system reads

\begin{equation}
\left| \Psi (t_1)\right\rangle =\aleph _2\prod_{i=1}^2\left\{ (1-p_i)D(\beta
)+(1+p_i)D(-\beta )\right\} \left| 1\right\rangle _1\left| 1\right\rangle
_2\left| \alpha \right\rangle  
\end{equation}

with $\aleph _2=\frac 14\left[ (1+\left| p_1\right| ^2)(1+\left| p_2\right|
^2)\right] ^{-1/2}$.We now set the internal state of the two ions to again
be in a similar superposition but with weighting factors $p_3$ and $p_4$ for
the two ions respectively, and also repeat this procedure $m$ cycles with
each cycle having the same interaction time $\Delta t$, then the motional
state of COM mode for the system of two ions is,

$$
\left| \Psi (t_m)\right\rangle =\aleph _{2m}\prod_{i=1}^{2m}\left\{
(1-p_i)D(\beta )+(1+p_i)D(-\beta )\right\} \left| \alpha \right\rangle 
$$
\begin{equation}
=\aleph _{2m}\sum_{k=0}^{2m}C_{2m}^kD\left[ 2(k-m)\beta \right] \left|
\alpha \right\rangle  
\end{equation}
where $\aleph _{2m}=\prod_{i=1}^{2m}\frac 12(1+\left| p_i\right| ^2)^{-1/2}$%
. and

\begin{equation}
C_{2m}^k=\sum_{\left\{ k\right\} }\prod_{i\in \left\{ k\right\}
}(1-p_i)\prod_{j\in \left\{ 2m-k\right\} }(1+p_j)  
\end{equation}

From above two equations we can see that if we let $2m=n$, then (15) and
(16) are the same with (8) and (9) respectively. In other words, the form of
the motional state by repeating $m$ cycles for two ions is the same as that
for one ion by repeating $2m$ cycles except for the difference between the
motional states for one ion and two ions. Therefore the parameters $p_i$ for
two ions by repeating $m$ cycles can be obtained from the parameters $p_i$
for one ion by repeating $2m$ cycles in terms of (11) and (12). thus we have
prepared a superposition of $2m$ coherent states on a line with required
coefficients for COM mode of two trapped ions by repeating $m$ cycles. Note
that this method can also be generated to more ions.

{\bf 3. The case for N ions.}

With $N$ ions confined in the linear trap, the motion of each ion will be
influenced by an overall harmonic potential due to the trap electrodes and
by the Coulomb force exerted by all of the other ions. They consist of a
multiple-ion-system. We usually describe this system in terms of COM mode
and other high-order vibrational modes. But now we only consider the
contribution from COM mode, and neglect for a while the effects from all
other high-order vibrational modes due to the large detunings with that
high-order sidebands. Now we prepare the initial state of the system of $N$
ions to be,

\begin{equation}
\left| \Psi (t_0)\right\rangle =\prod_{i=1}^N\frac 1{\sqrt{1+\left|
p_i\right| ^2}}\left( \left| 1\right\rangle _i+ip_i\left| 0\right\rangle
_i\right) \left| \alpha \right\rangle .  
\end{equation}
with parameters $p_i$ ($i=0,\cdots N$) are the weight factors.

Then illuminate all $N$ ions simultaneously with two lasers , i.e.,
performing the propagator (4) on the state (17), for an interaction time $%
\Delta t$, following by a measurement on the internal state of all $N$ ions.
If we detect all ions in excited state $\left| 1\right\rangle $, then the
conditioned motional state for the $N$ ions reads

$$
\left| \Psi (t_1)\right\rangle =\aleph _N\prod_{i=1}^N\left\{ (1-p_i)D(\beta
)+(1+p_i)D(-\beta )\right\} \left| \alpha \right\rangle  
$$
\begin{equation}
=\aleph _N\sum_{k=0}^NC_N^kD\left[ (2k-N)\beta \right] \left| \alpha
\right\rangle . 
\end{equation}
This
superposition form of coherent states for $N$ ions is completely the same as
(8) for one ion by repeating $n=N$ times, and the coefficients $\aleph _N$
and $C_N^k$ are also the same as (9) with $n=N$. But the implications for
these two cases are completely different: (8) represents a superposition of
COM vibrational mode for one ion, however (18) represents a superposition of
COM vibrational mode for $N$ ions. When $N$ is large we need only one cyclic
operation to obtain a superposition of a number of coherent states on a
line, which may need many cycles in the case for one ion. Owing to the
operations on each ion can be performed simultaneously, so it can saves much
time in process of state-preparation. It is important for the experimental
realization of quantum-state engineering in the view of decoherence. It is
also convenient for controlling the time that addressing lasers on each ion
to be the same, because of the dispersively illuminating on each ion for the two
lasers.

\begin{center}
{\bf IV. EFFECT OF SPECTATOR VIBRATIONAL MODES}
\end{center}

In previous section we neglect the presence of other vibrational modes
because of the large detunings from the sidebands of these spectator
vibrational modes and prepare the initial state as a superposition of
coherent states for COM mode. In this section we shall estimate the effect
of other spectator vibrational modes. We can see from the below discussions
that, on the conditions of Lamb-Dicke limit and RWA approximation, the COM
vibrational mode is isolated from all other vibrational modes in our quantum
state-preparation, so that the preparation of COM mode is not affected.

With $N$ ions in the trap which are strongly bounded in $y$ and $z$
directions but weakly bounded in a harmonic potential in $x$ direction, we
need only consider $N$ longitudinal modes. The Hamiltonian for this system
is[16]

\begin{equation}
H=H_0+H_{int},  
\end{equation}

$$
H_0=\sum_{l=1}^N\nu _l(a_l^{+}a_l+1/2)+\omega \sum_i^N\sigma _{iz}, 
$$

$$
H_{int}=\sum_i^N\frac \Omega 2\left\{ \sigma _{+i}\left[ e^{-i\omega
_1t}+e^{-i\omega _2t}\right] e^{i\sum_{l=1}^N\eta
_{i,l}(a_l+a_l^{+})}+H.C.\right\} 
$$
where $\nu _l$ and $a_l^{+}$ ($a_l$) are the frequency and ladder operators
of the $l$th mode. The excursion of the $i$th ion in the $l$th mode is
described by the Lamb-Dicke parameter $\eta _{i,l}$ which may be represented
as $\eta _{i,l}=\eta (\sqrt{N}b_i^l/\sqrt{\nu _l/\nu })$ , where $\eta $ and 
$\nu $ refer to COM mode as in the previous section, and where $b_i^l$ obeys
the orthogonality conditions $\sum_{i=1}^Nb_i^lb_i^{l^{^{\prime }}}=\delta
_{l,l^{^{\prime }}}$ and $\sum_{i=1}^Nb_i^lb_{i_{^{\prime }}}^l=\delta
_{i,i_{^{\prime }}}$[17]. For COM vibrational mode we have $l=1$ and $%
b_i^1=1/\sqrt{N}$ for all ions. And $l=2$ represents breathing mode which
corresponds to each ion oscillating with an amplitude proportional to its
equilibrium distance from the trap center, and $l>2$ denote high-order
normal modes. Similar to the derivation in Sec.II, in the Lamb-Dicke limit
and by use of RAW approximation, we find the Hamiltonian in the interaction
picture with respect to $H_0$ is,

\begin{equation}
H_{int}=\sum_{l=1}^N\Theta _l\left[ x_lf_l(t)+p_lg_l(t)\right] 
\end{equation}

where $f_l(t)=-2\sqrt{2}\eta \Omega \sqrt{\nu /\nu _l}\cos (\nu _l-\delta )t$
and $g_l(t)=-2\sqrt{2}\eta \Omega \sqrt{\nu /\nu _l}\sin (\nu _l-\delta )t$
, and the internal and motional state operators are defined by $\Theta _l=%
\sqrt{N}\sum_{i=1}^Nb_i^lj_{yi}$ and $x_l=(1/\sqrt{2})(a_l+a_l^{+})$ and $%
p_l=(i/\sqrt{2})(a_l^{+}-a_l)$ . Note that the ladder operators for
different modes commute and spin operators for different ions commute, so we
have the following propagator,

\begin{equation}
U=\prod_{l=1}^N\prod_{i=1}^N\frac 12\left\{ [D^{+}(\beta _i^l)+D(\beta
_i^l)]-\sigma _{yi}[D^{+}(\beta _i^l)-D(\beta _i^l)]\right\} 
\end{equation}

with $\beta _i^l=i\eta \Omega t\sqrt{N\nu /\nu _l}b_i^le^{i(\nu _l-\delta
)t} $ .

Now we assumed that the motional state for all modes are in ground and the
whole state for the system of $N$ ions is

\begin{equation}
\left| \Psi (t_0)\right\rangle =\prod_{i=1}^N\frac 1{\sqrt{1+\left|
p_i\right| ^2}}\left( \left| 1\right\rangle _i+ip_i\left| 0\right\rangle
_i\right) \prod_{l=1}^N\left| 0\right\rangle _l. 
\end{equation}
where we have assumed that all vibrational modes are in ground state.

After performing the propagator (21) on the initial state (22) for some
interaction time $\Delta t$, subsequent probing the internal state of all $N$
ions, with detection of no fluorescence for every ion, then the conditioned
motional state reads

\begin{equation}
\left| \Psi (t_1)\right\rangle =\prod_{l=1}^N\left\{ \prod_{i=1}^N\frac 1{2%
\sqrt{1+\left| p_i\right| ^2}}\left[ (1-p_i)D(\beta _i^l)+(1+p_i)D(-\beta
_i^l)\right] \left| 0\right\rangle _l\right\} . 
\end{equation}

From (23) we can see that the COM vibrational mode is isolated from all
other motional modes. In fact all vibrational modes are isolated from each
other. For COM vibrational mode ($l=1$), we have $\beta _i^1=i\eta \Omega
te^{i(\nu _1-\delta )t}=\beta $, thus we again acquired the same prepared
quantum-state as before, except for the difference of the initial displacing 
point. And the existence of other motional modes did not
affect the preparation of COM vibrational mode. For spectator vibrational
modes($l\geq 2$) we have $\nu _l\geq \sqrt{3}\nu $, so that $(\nu _l-\delta
)\gg (\nu -\delta )$ for the lasers very close to the first sidebands $%
\left( \delta \approx \nu \right) $. If we select the interaction time 
between each ion and lasers in each cycle satisfying 
$(\nu -\delta )t\sim 1 $, then the displacing factor $\beta _i^l$
oscillate very much times and the averages over time
disappear, thus the spectator vibrational modes only endured very small
effects. This is why we can neglect the contribution from these vibrational
modes in previous sections.

It is worthwhile to note that the above conclusions are obtained in
conditions of Lamb-Dicke limit and low-excitation regime, where the two
lasers performing on each ion are very close to the red and blue sidebands
for COM mode. If these conditions were not met, then the COM mode would be
entangled with other spectator vibrational modes. However these conditions
can be met well in practice, therefore the results concluded in this paper
are valid.

\begin{center}
{\bf V. CONCLUSION}
\end{center}

In this paper we have shown that how an arbitrary superposition of
equidistant coherent states can be created on a line in phase space for the
COM mode of $N$ ions. This approach can be viewed as the generation of the
previous methods for preparing motional states of one ion. After an initial
preparation of internal state of every ion, two lasers are projected on each
ion to entangle the vibrational modes with internal states. And then
measuring the internal state of each ion, with no fluorescence for all ions
one disentangles the vibrational modes and creates a superposition of ionic
motion. By performing this procedure several times according to ones
requirements, the desired superposition of displace initial coherent state
on a line is generated for the COM vibrational mode of $N$ ions.

We note that for large number of ions one needs only one cyclic operation to
obtain such a superposition of many coherent states for the COM vibrational
mode. Owing to the operations on each ion can be performed simultaneously,
thus the operation time for quantum-state preparation is greatly reduced. It
is important for the experimental realization in the view of decoherence. We
also note that in our method the resulting motional state of COM mode is
isolated from that of other spectator modes in the conditions of Lamb-Dicke
limit and low-excitation, and the lasers very close to the first sidebands.
These conditions can be met easily in practice. Thus we can successfully
prepare the motional state of COM mode to a certain superposition even
though the existence of other vibrational modes.

\end{document}